%% file: sample-sigconf.tex
\documentclass[preprint]{acmart}

\usepackage{booktabs} 

\usepackage{wrapfig}
\usepackage{times}
\usepackage{amsmath}
\usepackage{algorithm2e}
\usepackage{bm}
\usepackage{multirow}
\usepackage{tcolorbox}
\usepackage{eurosym}
\tcbuselibrary{theorems}
\newtcbtheorem[number within=section]{mydefinition}{Model}%
{colback=yellow!5,colframe=yellow!35!black,fonttitle=\bfseries}{th}
\newtcbtheorem[number within=section]{mydefinition2}{Definition}%
{colback=blue!5,colframe=blue!35!black,fonttitle=\bfseries}{th}
\newtcbtheorem[number within=section]{mydefinition3}{Rule of Thumb}%
{colback=red!5,colframe=red!35!black,fonttitle=\bfseries}{th}
\usepackage{helvet}
\usepackage{courier}
\usepackage{epsf}
\usepackage{epsfig}
\usepackage{mathrsfs}
\usepackage{subfigure}
\usepackage{graphics}
\usepackage{booktabs}
\usepackage{balance}
\usepackage{graphicx}
\usepackage{bbm}
\usepackage{balance}
\usepackage{epstopdf}
\usepackage{pifont}

\usepackage{amsfonts}

\usepackage{pifont}

\newcommand{\method}{{\tt DeepHoops}}
\newcommand{\twindow}{T}
\newcommand{\epv}{$\epsilon_{\pi}$}
\newcommand{\epa}{$\Delta$\epv}
\newcommand{\maction}{$\alpha$}
\newcommand{\none}{{\tt null}}
\newcommand{\pps}{\beta_{\pi}}
\newcommand{\taction}{e}

\usepackage{xcolor}
\usepackage{epstopdf}
\usepackage{etoolbox}
\usepackage{tcolorbox}
\usepackage{tabularx}
\usepackage{array}
\usepackage{colortbl}
\tcbuselibrary{skins}

\DeclareRobustCommand{\officialeuro}{%
  \ifmmode\expandafter\text\fi
  {\fontencoding{U}\fontfamily{eurosym}\selectfont e}}

\newcolumntype{Y}{>{\raggedleft\arraybackslash}X}

\tcbset{tab1/.style={fonttitle=\bfseries\large,fontupper=\footnotesize\sffamily,
colback=yellow!10!white,colframe=red!75!black,colbacktitle=maroon!40!white,
coltitle=black,center title,freelance,frame code={
\foreach \n in {north east,north west,south east,south west}
{\path [fill=red!75!black] (interior.\n) circle (3mm); };},}}

\tcbset{tab2/.style={enhanced,fonttitle=\bfseries,fontupper=\footnotesize\sffamily,
colback=yellow!10!white,colframe=red!50!black,colbacktitle=blue!40!white,
coltitle=black,center title}}

\tcbset{tab3/.style={enhanced,fonttitle=\bfseries,fontupper=\footnotesize\sffamily,
colback=yellow!10!white,colframe=red!50!black,colbacktitle=red!40!white,
coltitle=black,center title}}

\setcopyright{rightsretained}
\newcommand{\todo}[1]{{\color{red} #1 }}

\acmDOI{10.475/123_4}

\acmISBN{123-4567-24-567/08/06}

\acmConference[KDD'19]{ACM KDD conference}{August 2019}{Anchorage, Alaska, US}
\acmYear{2019}
\copyrightyear{2016}

\acmArticle{4}
\acmPrice{15.00}

\begin{document}
\title{\method: Evaluating Micro-Actions in Basketball Using Deep Feature Representations of Spatio-Temporal Data}
\author{Anthony Sicilia}
\affiliation{%
  \institution{School of Computing \& Information \\ University of Pittsburgh}
}

\author{Konstantinos Pelechrinis}
\affiliation{%
  \institution{School of Computing \& Information \\ University of Pittsburgh}
}

\author{Kirk Goldsberry}
\affiliation{%
  \institution{School of Business \\ University of Texas, Austin}
}

\renewcommand{\shortauthors}{A. Sicilia et al.}
\renewcommand{\shorttitle}{{\method}}

\begin{abstract}
How much is an on-ball screen worth? How much is a backdoor cut away from the ball worth? Basketball is one of a number of sports which, within the past decade, have seen an explosion in quantitative metrics and methods for evaluating players and teams. 
However, it is still challenging to evaluate individual off-ball events in terms of how they contribute to the success of a possession. 
In this study, we develop an end-to-end deep learning architecture ({\method}) to process a unique dataset composed of spatio-temporal tracking data from NBA games in order to generate a running stream of predictions on the expected points to be scored as a possession progresses. 
We frame the problem as a multi-class sequence classification problem in which our model estimates  probabilities of {\em terminal actions} taken by players (e.g. take field goal, turnover, foul etc.) at each moment of a possession based on a sequence of ball and player court locations preceding the said moment. 
Each of these terminal actions is associated with an expected point value, which is used to estimate the expected points to be scored. 
One of the challenges associated with this problem is the high imbalance in the action classes. 
To solve this problem, we parameterize a downsampling scheme for the training phase. 
We demonstrate that {\method} is well-calibrated, estimating accurately the probabilities of each terminal action and we further showcase the model's capability to evaluate individual actions (potentially off-ball) within a possession that are not captured by boxscore statistics.
\end{abstract}

%
%
\maketitle

\input{samplebody-conf}

\input{related}
\input{methods}
\input{results}

\input{discussion}

\bibliographystyle{ACM-Reference-Format}
\bibliography{sample-bibliography}
\balance

\end{document}

%% file: samplebody-conf.tex
\section{Introduction}
\label{sec:intro}

While data have been an integral part of sports since the first boxscore was recorded in a baseball game during the 1870s, it is only recently that machine learning has really penetrated the sports industry and has been utilized for facilitating the operations and decision making of sports franchises.  
One main reason for this is our current ability to collect more fine-grained data; data that capture (almost) everything that happens on the court/field. 
For instance, since the 2013-14 season, the National Basketball Association (NBA) has mandated its 30 teams to install an optical tracking system that collects information 25 times every second for the location of all the players on the court, as well as the location of the ball. 
These data are further annotated with other information such as the current score, the game and shot clock time etc. 
Optical tracking data provide a lens to the game that is much different from traditional player and team statistics. 
Many actions that can affect the outcome of a game/possession happen away from the ball (off-ball actions) and are not recorded in boxscore-like metrics that capture almost exclusively on-ball events. 
For instance, while boxscore statistics such as steals and blocks do not accurately reflect the defensive performance of a team, analyzing the spatial movement of players can provide better insights in the (individual and team) defensive ability \cite{franks2015characterizing}. 
As another example, the Toronto Raptors - who were among the first teams to make use of this technology - were able to identify optimal positions for the defenders, given the offensive formation \cite{grandland}.  

Micro-actions refer to individual basketball moves by players, that combined, create the team's offensive and/or defensive execution.  
A screen, a pass, or a backdoor cut are examples of such micro-actions. 
These micro-actions, although important for the successful execution of a team's game plan, are rarely evaluated. 
Even when we have volume statistics from these actions, the context is usually absent, which makes it hard to evaluate the importance of each of these actions. 
For example, simply counting passes can be very misleading; a pass exchange in the backcourt is certainly not as valuable as a pass that leads to an open three point shot. 
In order to achieve our goal of evaluating micro-actions such as a pass, we use a unique dataset from approximately 750 NBA games from the 2016-17 regular season that includes the aforementioned information captured by the optical tracking system. 
We build a deep learning framework, namely {\method}, that predicts the running probability distribution of {\bf terminal actions} conditional to the players on the court, as well as, the latter's and the ball's spatio-temporal sequence over a time window $W$ of length {\twindow}. 
Here, terminal actions refer to actions that lead to change or reset of ball possession. 
We consequently transform this distribution to an expected points value {\epv} that essentially captures the expected points to be scored from the offense during this possession. 
{\method} also obtains an embedding for the players that is fed into the classifier hence, accounting for the personnel on court. 
This is particularly important since the same offensive scheme could have very different outcomes depending on who (for the offense and the defense) is on the court. 

\begin{figure}[ht]
\centering
\includegraphics[scale=0.25]{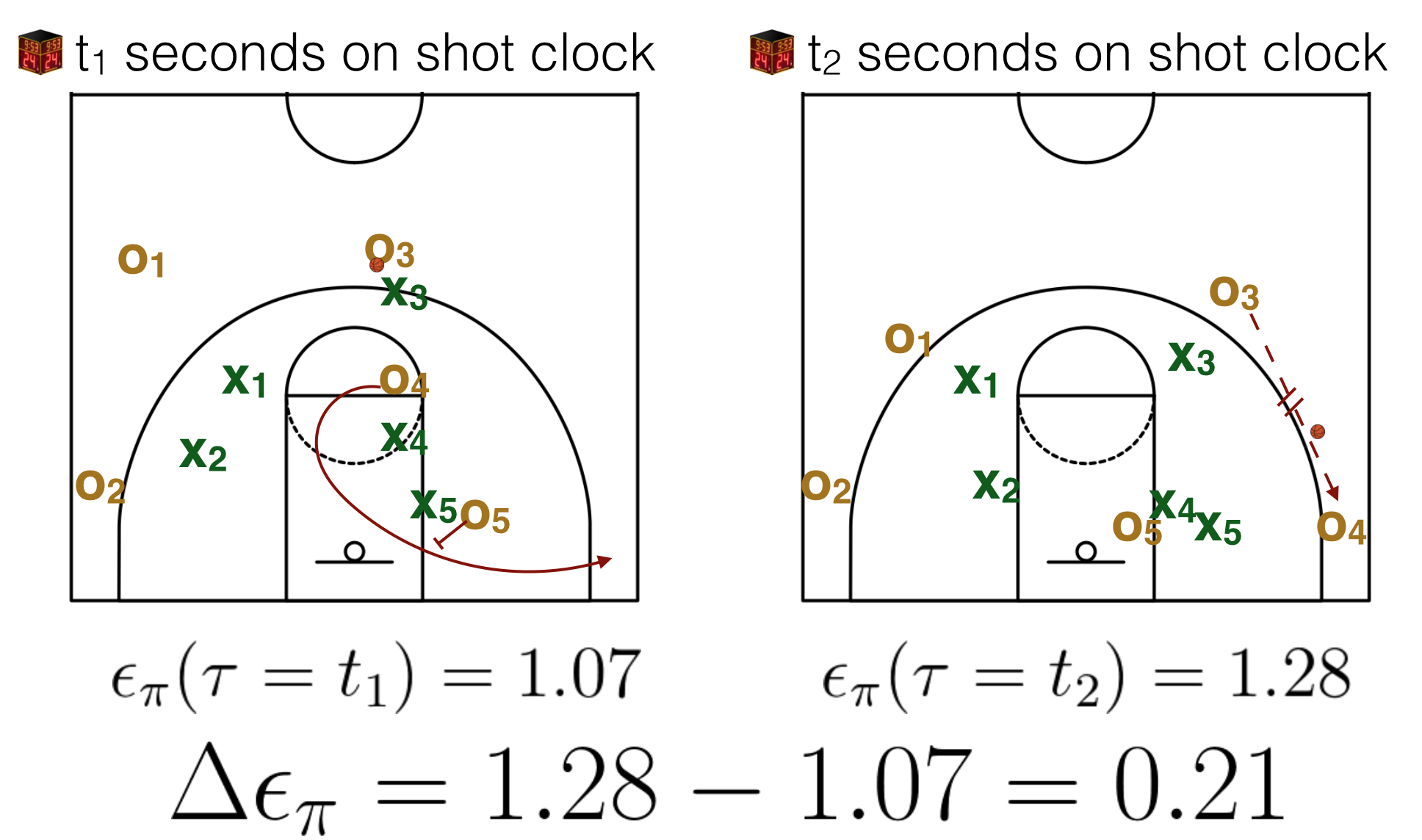}
\caption{{\method} is able to capture the expected points to be scored by the offense for a given possession snapshot. This allows us to estimate how micro-action(s) affect the scoring chances of a team. In the example above, the off-ball movement of player $O_4$ in conjunction with the screen by $O_5$ and the pass from $O_3$ increased the expected points from 1.07 to 1.28.}
\label{fig:deephoops}
\end{figure}

With {\method} providing an expected points value {\epv}$(\tau)$ at any time $\tau$ during the possession, we can calculate the expected points added by a micro-action {\maction} at time $\tau$ as {\epa}$($\maction$,\tau)$={\epv}$(\tau+\varepsilon)$-{\epv}$(\tau-\varepsilon)$. 
As an illustrative example, Figure \ref{fig:deephoops} presents two snapshots from an offensive set. A player of the offense ($O_4$) moves towards the left corner three area and receives a screen from $O_5$. 
He eventually gets the pass from the ball handler, $O_3$, and takes an open shot. 
During the first snapshot at time $t_1$, the expected points to be scored are {\epv}$(\tau=t_1)=1.08$. After the screen and pass for an open corner three, this increases to {\epv}$(\tau=t_2)=1.27$, giving a {\epa}$=0.21$.  
With this approach, we can start dissecting a possession to its smaller actions and quantify the overall impact of these actions to the expected final outcome. 
Furthermore, the applications of {\method} extend to ``what-if'' scenarios as well. 
What should we have expected to happen if $O_3$ did not pass to $O_4$ for the open shot, but drove to the basket since there is an open lane in the middle with $X_4$ and $X_5$ trying to close to the corner ? 
Will a different decision by the players on the court have led to higher expected points for the possession ? 
This can drive evaluations of on court decisions by the players. 
Similar evaluations can happen for the defense. 
What would be the value of {\epv}$(\tau=t_2)$ if $X_5$ had hedged the screen better ?  

{\bf Contributions: }
Our study contributes to the increasing literature on sports analytics by providing a generic deep learning framework that can quantify the value of micro-actions in basketball. 
As we will discuss in the following section, our work utilizes deep neural networks to essentially build on, and expand, the Markov chain model developed by Cervone {\em et al.} \cite{cervone2016multiresolution} to estimate the expected points value.
{\method} does not require any type of feature selection - e.g., definition and modeling of transitions between states - but rather obtains as input the raw trajectories of all the players and the ball. 
It consequently learns the most predictive features for the terminal action of a possession. 
Furthermore, it accounts for the players who are on the court through a multi-dimensional embedding. 
This architecture importantly allows {\method} to potentially be applied with minimal adjustments to any other {\em fluid} sport (e.g., soccer, American football, volleyball etc.). 
In particular, there is no need for defining any sport-specific features, states, or transitions between them. 



The rest of the paper is organized as follows: 
Section \ref{sec:related} discusses work relevant to our study and provides the required technical background, while Section \ref{sec:methods} presents the architecture of {\method}. 
Section \ref{sec:results} presents the evaluation of our method, while Section \ref{sec:discussion} concludes and discusses the scope and limitations of our study.

%% file: related.tex
\section{Background and Related Work}
\label{sec:related}
In this section we will provide some background on the neural network components we will include in the {\method} architecture. 
We will also discuss relevant studies on basketball analytics and more specifically on the use of spatio-temporal data in sports. 

\begin{table}[]
\begin{tabular}{lll}
\textbf{Notation/Term} & \textbf{Description}\\
\hline
{\epv} & Expected Points\\
{\epa} & Expected Points Added\\
 $\bm{x}_t^{(i)}$ & A moment during possession $i$ capturing the \\
 & locations of players and the ball at time $t$ \\
 $\bm{s}^{(i)}$ & Players on court during possession $i$ \\
$W_{\tau}^{(i)}$ & A temporal sequence window of moments \\
& during possession $i$ that ends at time $\tau$ \\
$T$ & Length of temporal window $W$\\
$r$ & Blind spot (buffer) at the end of window $W$ \\
$K$ & Downsampling parameter\\
{\maction} & Player micro-actions within play (pass, cut,\\ 
& screen etc.)\\
$\taction$ & Terminal action which ends a possesion (e.g.,\\
& shot attempt, turnover etc.)\\
$\mathcal{L}$ & Set of terminal actions \\
\end{tabular}
\caption{Notations used throughout the paper.}
\label{tab:notations}
\end{table}%

\subsection{Neural Networks}
\label{sec:networks}
There are primarily two neural network components included in the {\method} architecture: a Long-Short Term Memory Network (LSTM) and a Neural Embedding. 

\textbf{Long-Short Term Memory Network:} The LSTM network, originally introduced by Hochreiter and Schmidhuber \cite{hochreiter1997long} and later modified by Gers {\em et al.} \cite{gers1999learning} is a type of recurrent neural network (RNN) meant to solve the vanishing gradient problem \cite{Goodfellow-et-al-2016}. 
In general, these networks are capable of processing sequential data by using the current input as well as the previous state to compute the current state \cite{Goodfellow-et-al-2016}. 
For example, in the sports analytics domain, they have been used to process spatio-temporal data similar to ours for the task of offensive play-call classification \cite{wang2016classifying}. 
RNNs like other neural network models are thought to be able to represent complex functions more capably when designed with a deeper (multilayer) structure and there is empirical evidence to support this as well \cite{Goodfellow-et-al-2016, graves2013speech}. 
This multilayer structure can be accomplished by \textit{stacking} the output of each subsequent layer to form a stacked recurrent neural network (sRNN) \cite{pascanu2013construct}. 
We employ a stacked LSTM for the task of sequence modelling with specifics discussed in detail in Section \ref{sec:methods}.

\textbf{Neural Embedding:} When representing discrete objects as input to a neural network, it is often beneficial to embed the object in some latent space such that similar objects are {\em close} to each other in this space. 
For example, word embeddings have been widely used in the field of natural language processing for tasks such as sentence classification \cite{kim2014convolutional}. 
In the basketball setting, Wang and Zemel \cite{wang2016classifying} used an autoencoder with the shooting tendencies of a player as input in order to identify a latent representation for the players. 
This was then used to identify the position of a player based on his neighbors in the space. 
An embedding may also be learned during the training phase (with the other network parameters) as can be the case with word embedding \cite{kim2014convolutional}. 
This is the approach we take in {\method} where we employ a player embedding learned in tandem with the end-to-end architecture. 
In this way, two players will be close in the latent space based on their {\em impact} on the probability distribution of the terminal actions.  
We discuss this player embedding in detail in Section \ref{sec:methods}.
Finally, Table \ref{tab:notations} presents some of the mathematical notations we use in the rest of the paper. 

\subsection{Related Literature}
\label{sec:papers}

{\bf Player tracking data and basketball analytics:} The availability of optical tracking sports data has allowed researchers and practitioners in sports analytics to analyze and model aspects of the game that were not possible with traditional data. 
For instance, as alluded to above, Franks \textit{et al.} \cite{franks2015counterpoints} developed models for capturing the defensive ability of players based on the spatial information obtained from optical tracking data. 
Their approach is based on a combination of spatial point processes, matrix factorization and hierarchical regression models and can reveal information that cannot be inferred with boxscore data. 
As an example, the proposed model can identify whether a defender is effective because he reduces the quality of a shot or because he reduces the frequency of the shots all together. 
Furthermore, Miller \textit{et al.} \cite{miller14} use Non-Negative Matrix Factorization to reduce the dimensionality of the spatial profiles of player's shot charts. 
The authors use a log-Gaussian Cox point process to smooth the raw shooting charts, which they show provides more intuitive and interpretable patterns. 
Papalexakis {\em et al.} \cite{Papalexakis:2018:TMA:3269206.3272002} extended this approach using tensor factorization methods to build shooting profiles that can account for contextual information such as, shot clock, score differential etc. Additionally, using Bezier curves and Latent Dirichlet Allocation, a dictionary for trajectories that appear in basketball possessions was developed in \cite{miller2017possession}. 
In a different direction, Cervone \textit{et al.} \cite{cervone2016nba} computed the court's Voronoi diagram based on the players' locations and formalized an optimization problem that provides court realty values for different areas of the court. 
This further allowed the authors to develop new metrics for quantifying the spacing and the positioning of a lineup/team. 

Using optical tracking data, Yue \textit{et al.} \cite{yue2014learning} further developed a model with conditional random fields and non-negative matrix factorization for predicting the near-term actions of an offense (e.g., pass, shoot, dribble etc.) given its current state.  
In a tangential direction, D'Amour \textit{et al.} \cite{damour15} develop a continuous time Markov-chain to describe the discrete states a basketball possession goes through. 
Using this model the authors then propose entropy-based metrics over this Markov-chain to quantify the ball movement through the unpredictability of the offense, which also correlates well with the generation of opportunities for open shots.  
Seidl {\em et al.} \cite{seidlbhostgusters} further used optical tracking data to learn how a defense is likely to react to a specific offensive set using reinforcement learning. 
Recently, a volume of research has appeared that utilizes deep learning methods to analyze spatio-temporal basketball data, learn latent representations for players and/or teams, and identify and predict activities \cite{Mehrasa18,zhong18}, while additionally, Daly-Grafstein and Bornn \cite{daly2018rao} used the actual trajectory of a ball during a shot to obtain a robust estimation of the shooting skill of a player using a smaller sample of shots.

Close to our study is the work from Harmon {\em et al.} \cite{harmon2016predicting}, who utilized player trajectory and a convolutional neural network to predict whether a shot will be made or not. 
However, they only focus on possessions that end with a shot and they do not consider the development of the possession. 
The closest work to our study, as mentioned in Section \ref{sec:intro}, is that of Cervone \textit{et al.} \cite{cervone2016multiresolution} who utilized optical tracking data and developed a model for the expected possession value (EPV) using a multi-resolution stochastic process model. 
Our study builds on this work, providing a general framework which does not require the definition of features or possession states and can also be adjusted for applications in other sports. 
There are a few subtle but significant differences between {\method} and the work from Cervone \textit{et al.} \cite{cervone2016multiresolution}. 
In particular, in our work the expected points are calculated relative to a temporal window  instead of the possession as a whole. 
Hence, while \cite{cervone2016multiresolution} is more of a prediction model of what the offense will do next and how this will affect the end result of the possession, {\method} is more of a tool for value attribution given sequences that led to {\em good/bad} outcomes.
{\method} also includes a finer granularity for a possession's terminal actions (e.g., including classes for fouls, shooting fouls etc.). 
Overall, 

{\bf Other sports: } 
While basketball is the sport that has been studied the most through optical tracking data -- mainly due to the availability of data -- there is relevant literature studying other sports as well (both in terms of methodology and application). 
For example, Bialkowski \textit{et al.} \citep{bialkowski2014large} formulate an entropy minimization problem for identifying players' roles in soccer.  
They propose an EM-based scalable solution, which is able to identify the players' roles as well as the formation of the team using player tracking data. 
Lucey \textit{et al.} \citep{lucey2014quality} also used optical tracking data for predicting the probability of scoring a goal by extracting features that go beyond just the location and angle of the shot. 
More recently, Le \textit{et al.} \citep{Le2017CoordinatedMI} develop a collaboration model for multi-agents using a combination of unsupervised and imitation learning. 
They further apply their model to optical tracking data from soccer to identify the optimal positioning for defenders -- i.e., the one that \textit{minimizes} the probability of the offense scoring a goal -- given a particular formation of the offense.  
This allows teams to evaluate the defensive skills of individual players. 
In a tangential effort, Power \textit{et al.} \citep{Power:2017:PCE:3097983.3098051} define and use a supervised learning approach for the risk and reward of a specific pass in soccer using detailed spatio-temporal data. 
The risk/reward of a specific pass can further quantify offensive and defensive skills of players/teams. 
While we introduce {\method} as a framework for analyzing basketball data, it should be evident that it can really be used to analyze spatio-temporal (and in general multi-aspect) data for other sports as well. 

%% file: methods.tex
\section{Proposed Method}
\label{sec:methods}

In this section we formally define our problem and outline the details of our proposed deep learning solution. 
We begin with a discussion of the spatio-temporal tracking data, followed with the formal definition of our notion of expected points, which will be the medium through which we assign value to micro-actions. 
We then provide the details behind the {\method} architecture utilized to learn a deep feature representation of the spatio-temporal data, and consequently, estimate the expected points. 
Finally, we describe a technique used in the training phase to remedy the problem of highly unbalanced training data.

\subsection{Description and Processing of Spatio- Temporal Data}
\label{sec:data}
As aforementioned, to build {\method} we use a unique dataset composed of 750 NBA games from the 2016-17 regular season. 
Of primary interest is the optical tracking data which represents the NBA court as a three dimensional coordinate system in which the court location of offensive and defensive players can be expressed using rectangular coordinates along the $(x,y)$-plane and the court location and height of the ball can be expressed by taking into account the third dimension as well. 
During each second of gameplay, this information is recorded 25 times at evenly spaced intervals. 

The data is extensively annotated allowing for labeling of specific events and possession outcomes. 
Using this annotation it is simple to segment the game into well-defined team offensive possessions (i.e. into segments during which one team maintains control of the ball with the intention of scoring). 
Additionally, the data provide information about the specific players that are on the court as well as how much time remains on the shot-clock (which regulates the remaining maximum length of the possession). 

\textbf{Possessions:} Our dataset consists of a sample of more than 134,000 team possessions of interest. We define a possession $i$ as a sequence of $n$ \textbf{moments} $(\bm{x}_t^{(i)})_{t = 1}^{n}$ where each moment is a 24-dimensional vector, i.e., $\bm{x}_t^{(i)} \in \mathbb{R}^{24}$. 
The first 20 elements capture the court location of the 10 players via $(x,y)$-coordinates, the next three represent the court location and height of the ball via $(x,y,z)$-coordinates, while the last element represents the current value of the shot-clock. 
Each of these moments are well-annotated with event descriptions when relevant (e.g., a pass occurred during moment ($\bm{x}_{\tau}^{(i)})$). 
Additionally, each possession $i$ maintains information about the 5 offensive players $\{\bm{s}_j^{(i)}\}_{j=1}^{5}$ and 5 defensive players $\{\bm{s}_j^{(i)}\}_{j=6}^{10}$ on the court via one-hot encoding. 
Furthermore, we convert the court locations into polar coordinates with respect to the current offensive team's basket, since this way we are also encoding the {\em target} location and are likely more informative. 

\textbf{Temporal Window: } 
To estimate the expected points for a possession, we define a temporal window to act as the spatio-temporal input to {\method}. 
Specifically, for possession $i$ and moment $\tau$, we define a window of length {\twindow} 
as the subsequence: 

\begin{equation}
W_\tau^{(i)} = (\bm{x}_t^{(i)}) \quad\text{with}\quad \{\tau - ({\twindow} + r) \leq t \leq \tau - r\}
\label{eq:window}
\end{equation}
A window defined at moment $\tau$ captures {\twindow} moments of a possession leading up to the time of interest (i.e., $\tau$) with a buffer or blind spot $r$. 
As we will describe in the following, each of these windows will be assigned an outcome label $y_\tau^{(i)}$, which corresponds to a possession terminal action. 
The blind spot intuitively serves the purpose of pruning information late in the window, which are trivially indicative of a terminal action, i.e., defining the action (e.g., a shot that was just taken will have a larger value for the height of the ball). 

Furthermore, restricting to smaller (sliding) windows, instead of using the entirety of the trajectories from the beginning of the possession makes more sense in the basketball setting  \cite{harmon2016predicting}. 
Most plays develop over a small period of time (often less than 5 seconds) and if they are not successfully executed (e.g., finding an open shot) the offense resets its scheme. 
Hence, multiple plays may develop and fail during the same possession, so the temporal window effectively restricts the sample to only the information of primary interest preceding a given moment. 
For this reason, unless otherwise noted we will use $\text{\twindow}=128$ moments (just over 5 seconds\footnote{Recall that the temporal distance between two moments is equal to the sampling rate of the data, which is 0.04 seconds.}) with $r = 16$ moments (just over half of a second).

\textbf{Outcome Labels: }
As aforementioned, each window is labeled with an outcome $y_\tau^{(i)}$ that represents the type of (possession) terminal action that occurred at the end of the window. 
If the terminal action of the possession (e.g. a field goal attempt, a turnover etc.) occurs at moment $\tau$, then $W_\tau^{(i)}$ is labeled with this action, otherwise it is labeled as {\none}. 
The set of labels $\mathcal{L}$ consists of: 
\begin{itemize}
    \item \textbf{Field Goal Attempt}: The window ends with a field goal (FG) attempt, made or missed. 
    \item \textbf{Shooting Foul}: The window ends with a foul (illegal defensive action) during a shot attempt that awards free throw attempts to the player who was fouled.
    \item \textbf{Non-Shooting Foul}: The window ends with a player committing a non-shooting foul. 
    \item \textbf{Turnover}: The window ends with the offense committing a turnover (e.g., out-of-bounds pass, steal by the defense, offensive foul etc.), giving up possession of the ball to the opponent. 
    \item \textbf{\none}: No terminal action was recorded at the end of the window, and hence, the possession is still in progress. 
\end{itemize}


The inclusion of a {\none} label is crucial. 
As alluded to above an offensive set is usually executed over a short time period and then resets if needed. 
Therefore, labeling a whole possession, instead of sliding windows, would result in extensive noise in the training set which masks vital information. 
For example, let us consider a possession where the offense initially set up a screen that was ineffective. 
The offense had to reset, and after executing a different scheme, they took (and made) a corner three-point shot. 
If we only had a single label for this possession, then the ineffective screen action would be labeled with a terminal action that adds value (i.e., a FG attempt). 
However, with our labeling scheme - i.e., the use of the sliding temporal window - the ineffective screen will be included within a window labeled as {\none}, as it should, since it did not lead to an immediate shot/score or turnover.  
With this said, the label space $\mathcal{L}$ exhibits a very high imbalance, since the majority of the windows end with no event. 
The {\none} class out-weights all other class labels roughly 600 to 1. 
This class imbalance requires careful attention in the training phase. 
To handle this, we parameterize a downsample of the majority class (details discussed in Section \ref{sec:inner_method}). 

\subsection{Translating Labels to Points Per Possession}
\label{sec:ep}

Given a possession $i$ we can begin to define (a) the expected points {\epv}$^{(i)}(\tau)$ computed at each moment $\tau$ within the possession, as well as, (b) its application to the valuation of micro-actions within possessions.

\begin{figure}[ht]
\centering
\includegraphics[scale=0.3]{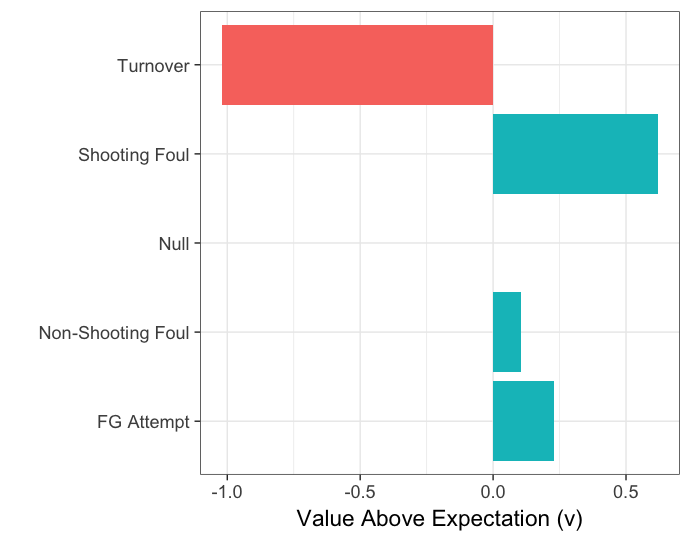}
\caption{For each outcome, we calculate the points above $\pps = 1.02$ the offense will gain if this outcome is realized.}
\label{fig:nu}
\end{figure}

\textbf{Expected Points:} Every possession has a baseline expected points $\pps$ to be scored. 
We can calculate this by dividing the total number of points scored over all games with the number of total possessions. 
In our dataset, $\pps = 1.02$. 
Using {\method} we can further adjust this value - in real time - based on the probability of the labels in $\mathcal{L}$. 
In particular, we define a function $\nu: \mathcal{L}\rightarrow\mathbb{R}$ that maps every class/terminal action to a number that captures the points above expectation (i.e., above $\pps$) to be scored if this terminal action is realized during the play. 
For example, the average points scored during a possession that terminates with a shot (average points per shot) is $1.25$. Hence, a possession that ends in a FG attempt is worth $1.25-\pps = 1.25-1.02=0.23$ points \textit{above} expectation, and therefore, we define $\nu(\text{FG~Attempt})=0.23$. 
Figure \ref{fig:nu} presents the mapping for every terminal action to points above $\pps$. 
For example, a turnover that ends in a change of possession has a value of $\nu = -\pps$ (because this terminal action counteracts any chance of scoring), while the {\none} class has a value of 0 (because this terminal action provides no new information with which to adjust the baseline). 
For the 
non-shooting fouls, we have incorporated the fraction of fouls committed while the defense is in the penalty (because in this situation free throws are awarded). 

The expected points $\text{\epv}^{(i)}(\tau)$ for possession $i$ at time $\tau$ can now be defined as an expectation above $\pps$ conditional to (i) the temporal window $W_\tau^{(i)}$, and (ii) the players on the court. 
Specifically, if $\text{y}$ is the discrete random variable that captures the outcome of window $W_\tau^{(i)}$ and follows the distribution $P(\text{y})$, then $\text{\epv}^{(i)}(\tau)$ is defined as: 
\begin{equation}%
    \text{\epv}^{(i)}(\tau) = \pps + \mathbb{E}_{\text{y} \sim P(\text{y})}[\nu(y) | W_\tau^{(i)}, \{\bm{s}_j^{(i)}\}_{j=1}^{10}]
    \label{eqn:expectedpoints}
\end{equation}%
For example, in the extreme (and rather unrealistic) case where all labels have a zero probability except the turnover that has a 100\% probability, the expected points will be equal to: $ \text{\epv}^{(i)}(\tau) = 1.02 - 1\cdot1.02 = 0$,  
as it should be since the team will not score and the ball will change possession. 

\textbf{Expected Points Added:} 
The primary goal of defining {\epv} is to assign a value to specific micro-action {\maction} as demonstrated in Figure \ref{fig:deephoops}. 
Intuitively, the inclusion of a valuable micro-action within a temporal window $W_\tau^{(i)}$ during possession $i$ should alter the distribution of outcomes, making higher value outcomes more likely conditional to the new window. Formally, if {\maction} occurs at moment $\tau$ within a possession $i$, the expected point added can be calculated as: 
\begin{equation}
    \text\epa(\text\maction,\tau)^{(i)}={\text\epv}^{(i)}(\tau+\varepsilon)-\text\epv^{(i)}(\tau-\varepsilon)
    \label{eqn:expectedpointsadded}
\end{equation}
where $\varepsilon > r$ to ensure that the micro-action {\maction} occurs in the new window $W_{\tau + \varepsilon}^{(i)}$. Notice, that this definition, in conjunction with Equation \ref{eq:window}, imply that taking $\varepsilon = \delta + r$ corresponds to a shift around the micro-action of $\delta$ moments.

\subsection{The {\method} Architecture}
\label{sec:inner_method}

\begin{figure}[ht]
\centering
\includegraphics[width=\linewidth]{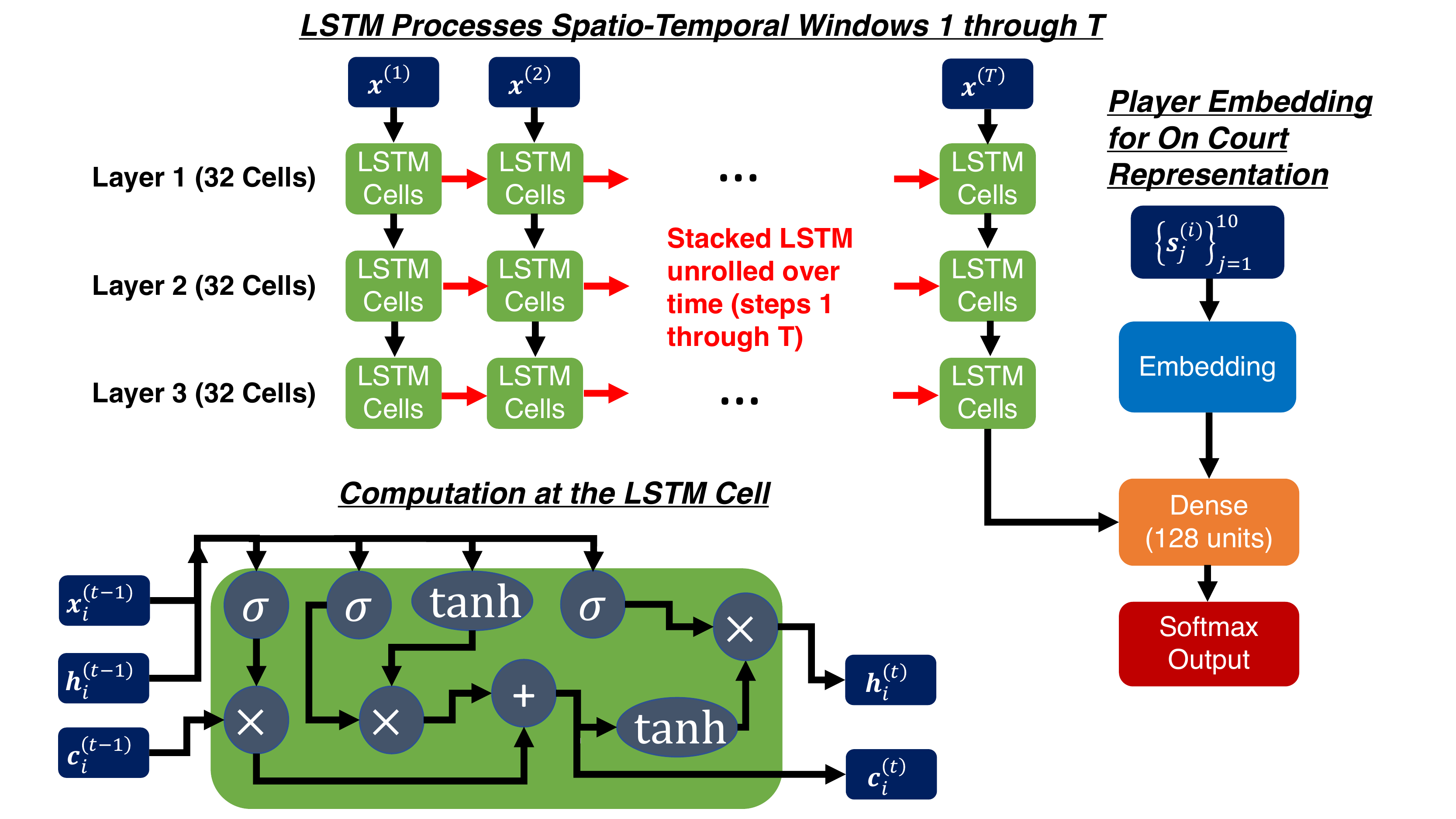}
\caption{Diagram of {\method}. The LSTM network (displayed unrolled) learns a feature representation of the spatio-temporal window. A latent space representation of on court players is concatenated and processed by an additional dense layer before softmax output for probability estimation. Lower left displays computation at the LSTM Cell.}
\label{fig:archdiagram}
\vspace{-0.15in}
\end{figure}

Based on the discussion above, the problem of calculating expected points and expected points added translates to estimating the probability distribution: 
$P(y | W_\tau^{(i)}, \{\bm{s}_j^{(i)}\}_{j=1}^{10})$. 
We model the probability of each label with a softmax function, that is,we estimate the probability for label $y_i$ as:
\begin{equation}
P(y | W_\tau^{(i)}, \{\bm{s}_j^{(i)}\}_{j=1}^{10})=\frac{e^{z_{i}}}{\sum_{y_j \in \mathcal{L}}e^{z_{j}}}
\label{eq:softmax}
\end{equation}
where $z_{i}$ is computed using the learned feature representation provided by the neural network $g$ with parameters $\bm\Theta$:
\begin{equation}
\bm{z} = \bm{W}^{\mathrm{T}}g(W_\tau^{(i)}, \{\bm{s}_j^{(i)}\}_{j=1}^{10}; \bm{\Theta}) + \bm{b}
\label{eq:nn}
\end{equation}
Here, $\bm{W}$ and $\bm{b}$ correspond to the linear layer producing the log outcome probabilities, and $\bm\Theta$ are the parameters of $g$ optimized to learn a joint feature representation of $W_\tau^{(i)}$ and $\{\bm{s}_j^{(i)}\}_{j=1}^{10}$ such that the negative log-likelihood is minimized \cite{Goodfellow-et-al-2016}. Formally, in expectation over the data distribution, that is: 
\begin{equation}
 \min_{\bm{W}, \bm{b},\bm{\Theta}}-\mathbb{E}\sum_{y_j \in \mathcal{L}}\mathbbm{1}\{y_j\}\log P(y_j | W_\tau^{(i)}, \{\bm{s}_j^{(i)}\}_{j=1}^{10}) 
 \label{eqn:first_loss}
\end{equation}
where $\mathbbm{1}$ is an indicator function. The remainder of this section is devoted to describing the network $g$ (Figure \ref{fig:archdiagram}) that forms the core component of {\method}. 

To jointly learn a feature representation of $W_\tau^{(i)}$ and  $\{\bm{s}_j^{(i)}\}_{j=1}^{10}$, we construct $g$ with two main modules. 
The first is a sequence processing module and uses a stacked  LSTM network to learn a representation of the spatio-temporal window leading up to time $\tau$. 
This module allows for important information about on-court actions (either on or off-ball) to be captured as the play progresses. 
The second module primarily serves the purpose of capturing information with regards to who is on the court. 
This additional information is meant to model the impact of the specific lineup playing on the expected value of the possession as discussed in Section \ref{sec:intro}. 
The output of the two main components is combined and fed into a dense (i.e. fully connected) layer to generate the final joint feature representation.

\textbf{Stacked LSTM Network:} 
The LSTM network, and in particular the deep LSTM, is a common, effective neural network for processing sequential inputs \cite{Goodfellow-et-al-2016, graves2013speech}. Each layer of the LSTM has a number of LSTM cells (in our implementation, we have 32 for each of the 3 layers). 
The state of the $i^{th}$ cell at timestep $t$ is represented by the $i^{th}$ element of vector $\bm{h}^{(t)}$. 
Here, the timesteps run over the length of our sequential data. 
Each of these cells processes data similarly to the standard unit of an RNN, that is, they compute an external state (i.e. $\bm{h}^{(t)}$) via information from previous states and the current input. 
However, there is an additional internal state $\bm{c}^{(t)}$ and a number of gates that control the flow of information \cite{Goodfellow-et-al-2016} (see Figure \ref{fig:archdiagram}). 
In our implementation \cite{chollet2015keras}, given a window, which we rewrite for clarity as $W_\tau = (\bm{x}^{(1)}, \ldots \bm{x}^{(\text\twindow)})$, the computation for cell $i$ at time step $t$ can be written as:

\begin{equation}
\bm{i}_i^{(t)} = \sigma(\bm{b}_i^\prime + \sum_{j}\bm{U}_{i,j}^\prime\bm{x}_{j}^{(t)} + \sum_{j}\bm{W}_{i,j}^\prime\bm{h}_j^{(t-1)})
\label{eq:1}
\end{equation}

\begin{equation}
\bm{f}_i^{(t)} = \sigma(\bm{b}_i^f + \sum_{j}\bm{U}_{i,j}^f\bm{x}_{j}^{(t)} + \sum_{j}\bm{W}_{i,j}^{f}\bm{h}_j^{(t-1)})
\label{eq:2}
\end{equation}

\begin{equation}
\bm{c}_i^{(t)} = \bm{f}_i^{(t)}\bm{c}_i^{(t-1)} + \bm{i}_i^{(t)}\tanh(\bm{b}_i^c + \sum_{j}\bm{U}_{i,j}^c\bm{x}_{j}^{(t)} + \sum_{j}\bm{W}_{i,j}^c\bm{h}_j^{(t-1)})
\label{eq:3}
\end{equation}

\begin{equation}
\bm{o}_i^{(t)} = \sigma(\bm{b}_i^o + \sum_{j}\bm{U}_{i,j}^o\bm{x}_{j}^{(t)} + \sum_{j}\bm{W}_{i,j}^o\bm{h}_j^{(t-1)})
\label{eq:4}
\end{equation}

\begin{equation}
\bm{h}_i^{(t)} = \bm{o}_i^{(t)}\tanh(\bm{c}_i^{(t)})
\label{eq:5}
\end{equation}
where in the above all subscripts are component indices and the $\bm{U}^{*}, \bm{W}^{*}, \bm{b}^{*}$ are the parameters to be learned. 
Conceptually, $\bm{i}$ is the input gate, $\bm{f}$ is the forget gate, and $\bm{o}$ is the output gate. 
Each of these gates controls what information is carried to the next time step (via internal state $\bm{c}$) and what information is output (via external state $\bm{h}$) \cite{Goodfellow-et-al-2016,gers1999learning}. As mentioned {\method} uses a (deep) stacked LSTM \cite{pascanu2013construct, graves2013speech}. 
This multilayer structure is formed by stacking the external state; i.e., any subsequent layer in the stacked LSTM takes $\bm{x}^{(t)}$ to be replaced with $\bm{h}^{(t)}$ of the previous layer. Hence, the stacked LSTM simply treats the external state of the previous layer as input to the next. Our stacked LSTM is restricted to processing fixed-length sequences (i.e. our windows of length {\twindow}), so the final state of a given layer is referenced as $\bm{h}^{(\text{\twindow})}$. We therefore define the output of our LSTM network as the final state $\bm{h}^{(\text{\twindow})}$ of the last layer which will be concatenated with the player representations discussed below before processing by a final dense layer as depicted in Figure \ref{fig:archdiagram}.

\textbf{Player Representation: }
As aforementioned, we jointly learn an embedding of individual players within our end-to-end architecture. 
The players are represented through vectors that are ``updated'' with each training sample.
This update, is formally dependent on its contribution to the output of the network $g$. Namely, the embedding implementation \cite{chollet2015keras} randomly initializes a group of dense $d$ dimensional vectors $\bm{a}^{j} \in \mathbb{R}^d$ to represent each player. If we form the tensor $\bm{A} = [\bm{a}^1, \ldots, \bm{a}^m] \in \mathbb{R}^{d \times m}$, then the one-hot encoding of the $j^\text{th}$ player $\bm{s}_j^{(i)}$ can be used to extract that player's representation via the matrix multiplication $\bm{A}\bm{s}_j^{(i)}$. As discussed, the extracted player representations are concatenated to the final state $\bm{h}^{(\text{\twindow})}$ of the last layer of the LSTM and fed through an additional dense layer before classification. 
This way, the player representations are jointly learned in the sense that whenever a player is present on court, his randomly initialized vector is updated during backpropogation to minimize the loss (Equation \ref{eqn:first_loss}) for the classification task (just as all other weights within the network $g$). 
Players with {\em similar} contributions to the distribution of terminal actions are then expected to be close in the corresponding latent space. 

\textbf{Downsampling Scheme: }
With the network $g$ defined, we now address the imbalance of the {\none} label. 
As mentioned in Section \ref{sec:data}, these {\none} temporal windows capture important information about micro-actions, but training on such highly imbalanced data could pose problems in generating well calibrated probabilities. 
One technique when dealing with high class imbalance is to down-sample the majority class \cite{chawla2009data}. 
We propose and use a problem-specific down-sampling technique which, importantly, ensures that each {\em possession} has equal representation in the newly generated training set. In particular, from a possession $i$, we uniformly sample $K$ time-steps $\{t_{i,k}\}_{k=1}^{K}$ for which the corresponding window $W_{t_{i,k}}^{(i)}$ is labeled as {\none}. 
Then, besides the single window $W_\tau^{(i)}$ which is labeled by this possession's terminal action, all other windows are discarded.
Hence, during the training phase, we compute the loss $L$ for a batch of size $N$ possessions with a down-sample of size $K$ as: 
\begin{equation}
    L_i = \sum_{y_j \in \mathcal{L}}\mathbbm{1}\{y_j\}\log P(y_j | W_\tau^{(i)}, \{\bm{s}_j^{(i)}\}_{j=1}^{10})
\end{equation}
\begin{equation}
    L_i^\prime = \sum_{k=1}^{K}\sum_{y_j \in \mathcal{L}}\mathbbm{1}\{y_j\}\log P(y_j | W_{t_{i,k}}^{(i)}, \{\bm{s}_j^{(i)}\}_{j=1}^{10})  
\end{equation}
\begin{equation}
    L = -\frac{1}{N(K+1)} \sum_{i = 1}^{N} L_i + L_i^\prime
\end{equation}

This technique is naturally parameterized by $K$ and we explore the impact of this parameter in Section \ref{sec:results}. 
We additionally re-generate the down-sampled dataset on each epoch (i.e. the sample $\{t_{i,k}\}_{k=1}^{K}$ is redrawn for each possession on any new epoch).

%% file: results.tex
\section{Experiments}
\label{sec:results}



\textbf{Experimental Setup: }
For our evaluations, we break the data (Section \ref{sec:data}) into training, validation, and test sets at a 75-10-15 split. 
All results presented are on the test set, while the 
the performance on the validation set is monitored in training to prevent overfitting via Early Stopping \cite{prechelt1998early}. 
The monitored metric is the Brier score (discussed in what follows) with a minimum required improvement of 0.01 over 5 epochs. 
Only the {\em best} performing model on the validation set is used in the testing process. 
For all experiments, we used the following hyperparameters for the {\method} architecture: 3 LSTM layers with 32 cells each, an embedding of dimensionality $d=8$, and 1 dense layer with 128 units prior to classification. To prevent overfitting \cite{srivastava2014dropout}, dropout with rate of 0.3 is applied to the output of the dense layer and variational dropout \cite{gal2016theoretically, chollet2015keras} is applied throughout the LSTM network with rate of 0.2 for transformation of both input and recurrent state. 

\subsection{Probability Calibration}
\label{sec:cali}

In the following, our main objective is to evaluate the accuracy of the predicted probabilities for the possible possession outcomes as this is directly indicative of accuracy of the expected points.
Evaluation of the quality of a probability, i.e., probability calibration, has traditionally been done through the Brier score and the reliability curves. 

{\bf Brier Score:} 
The Brier score is a proper scoring rule that quantifies the calibration of a set of probabilistic predictions \cite{brier1950verification, murphy1973hedging,gneiting2007strictly}. 
If we have $N$ samples (i.e., predictions) and $R$ possible outcomes, the Brier score is calculated as:
\begin{equation}
BS = \frac{1}{N}\sum_{i = 1}^{N} \sum_{r = 1}^{R}(f_{i,r} - o_{i,r})^2 
    \label{eqn:brier}
\end{equation}
where $f_{i,r}$ is the model predicted probability for outcome/label $r$ on sample $i$ and $o_{i,r}$ is 1 if sample $i$'s label is $r$, while it is $0$ otherwise. 
From the above definition it should be evident that between two models, the one with the lowest Brier score is better calibrated. 

Typically the Brier score of a model is compared to a climatology model which assigns to every outcome its baseline probability, i.e., the historical frequency of occurrences for each outcome \cite{brier1950verification, murphy1973hedging}. 
Using the Brier score of this climatology model $BS_{ref}$, we can further calculate the \textit{skill} of probability estimates using the Brier Skill Score ($BSS$) \cite{murphy1973hedging, gneiting2007strictly}. 
Specifically, if $BS_{\text{ref}}$ is the Brier score of the climatology model, then $BSS$ is calculated as:  
\begin{equation}
    BSS = 1 - \frac{BS}{BS_{\text{ref}}} 
    \label{eqn:brier_skill}
\end{equation}
$BSS$ will be equal to 1 for a model with perfect calibration, i.e., $BS = 0$. 
A model with no {\em skill} over the climatology model, will have a value of 0 since $BS = BS_{\text{ref}}$. 
If $BSS < 0$, then the model exhibits less skill than even the reference model. 

Table \ref{tab:brier} displays the Brier score of {\method} and the corresponding climatology model, while additionally showing the Brier Skill Score of {\method} for different values of $K$ (the downsampling rate). 
We also include the minimum training time\footnote{All models were trained on an iMac with 3.3GHz Intel Core i7 processor (GPU not used).} for an epoch  of {\method} for different values of $K$. 
As can be seen, regardless of the value of $K$, {\method} exhibits better calibration as compared to that of the climatology model. Additionally, by comparing the Brier Skill Score, we see that improvement over the climatology model is best when $K=2$. 
Furthermore, as one might have expected, higher value of $K$ requires longer time to train. 
For this reason, in the remainder of the experiments, we pick $K=2$. 

\begin{figure}[ht]
\centering
\includegraphics[scale = 0.6]{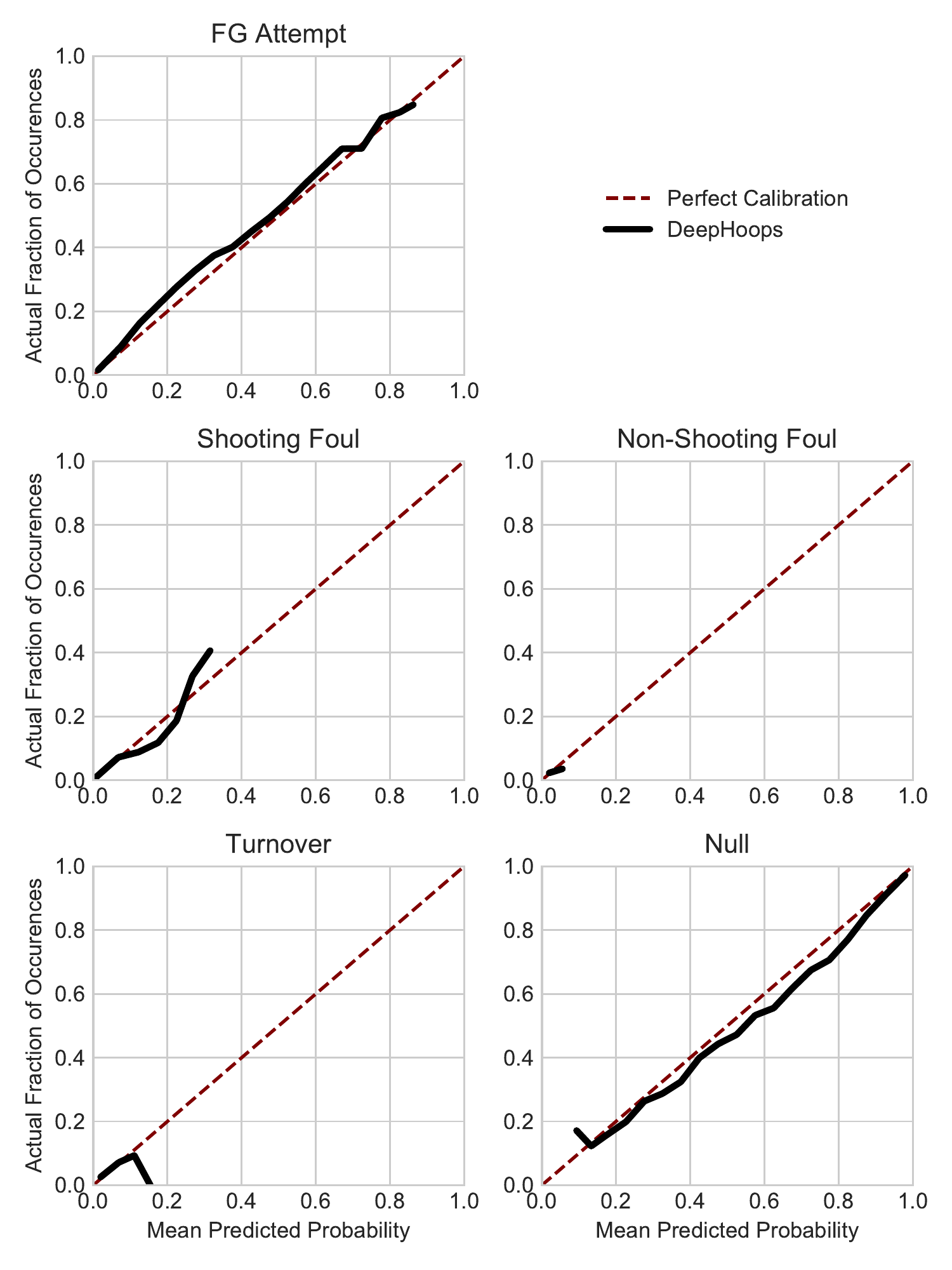}
\caption{Reliability Curves for {\method}' probability estimates. The dashed line $y=x$ represents perfect calibration. {\method} follows this line closely, estimating accurately the outcome probabilities. The number of bins is 20. The tail in the turnover class is due to a bin containing only 1 estimate.}
\label{fig:reliability}
\end{figure}

\textbf{Reliability Curves:} 
In order to estimate the accuracy of the output probabilities for each label, ideally we would like the same window within a possession to {\em replay} several times and then estimate the number of times that it ended up to each possible outcome. 
However, obviously this is not possible and hence, we will rely on the reliability curves. 
In particular, if the predicted probabilities were accurate, when considering all the windows  where terminal action $\taction$ was predicted  with a probability of x\%, then terminal action $\taction$ should have been observed in (approximately) x\% of these accidents. 
Given the continuous nature of probabilities, for a particular outcome class, the estimated probabilities for this class are binned (in this case by intervals of size 0.05). 
For each bin, the fraction of actual occurrences of this class is then calculated. 
Therefore, perfect calibration occurs when the fraction of occurrences is equal to the estimated probabilities (i.e. along the line $y = x$). Figure \ref{fig:reliability} provides reliability curves for the predicted probabilities of {\method}. Overall, {\method} is shown to be well-calibrated. In addition to low-probability events, outcome classes such as FG Attempt, Shooting Foul, and {\none} with a large range or probabilities, are accurately estimated.
\begin{table}[]
\begin{tabular}{l|llll}
    & $BS$ & $BS_{\text{ref}}$ & $BSS$ & Epoch Time ($s$)\\
    \hline
$K=1$ & $0.4569$   & $0.6070$   & $0.2472$ &   2180 \\ 
$K=2$ & $0.3598$   & $0.4920$   & $\textbf{0.2686}$ &   2929 \\
$K=3$ & $0.3094$   & $0.4017$   & $0.2299$    & 3552 \\
$K=4$ & $0.2659$   & $0.3371$   & $0.2114$ & 4200\\
\hline
\end{tabular}%
\caption{{\method} Brier Score ($BS$), Climatology Model Brier Score ($BS_{\mathrm{ref}}$), and {\method} Brier Skill Score ($BSS$). {\method} outperforms the climatology (baseline) model in all cases. Performance is best for $K=2$ (among the values examined). Epoch Time (in seconds) is lowest over all epochs.}
\label{tab:brier}
\end{table}

\begin{figure}[ht]
\centering
\includegraphics[scale=0.4]{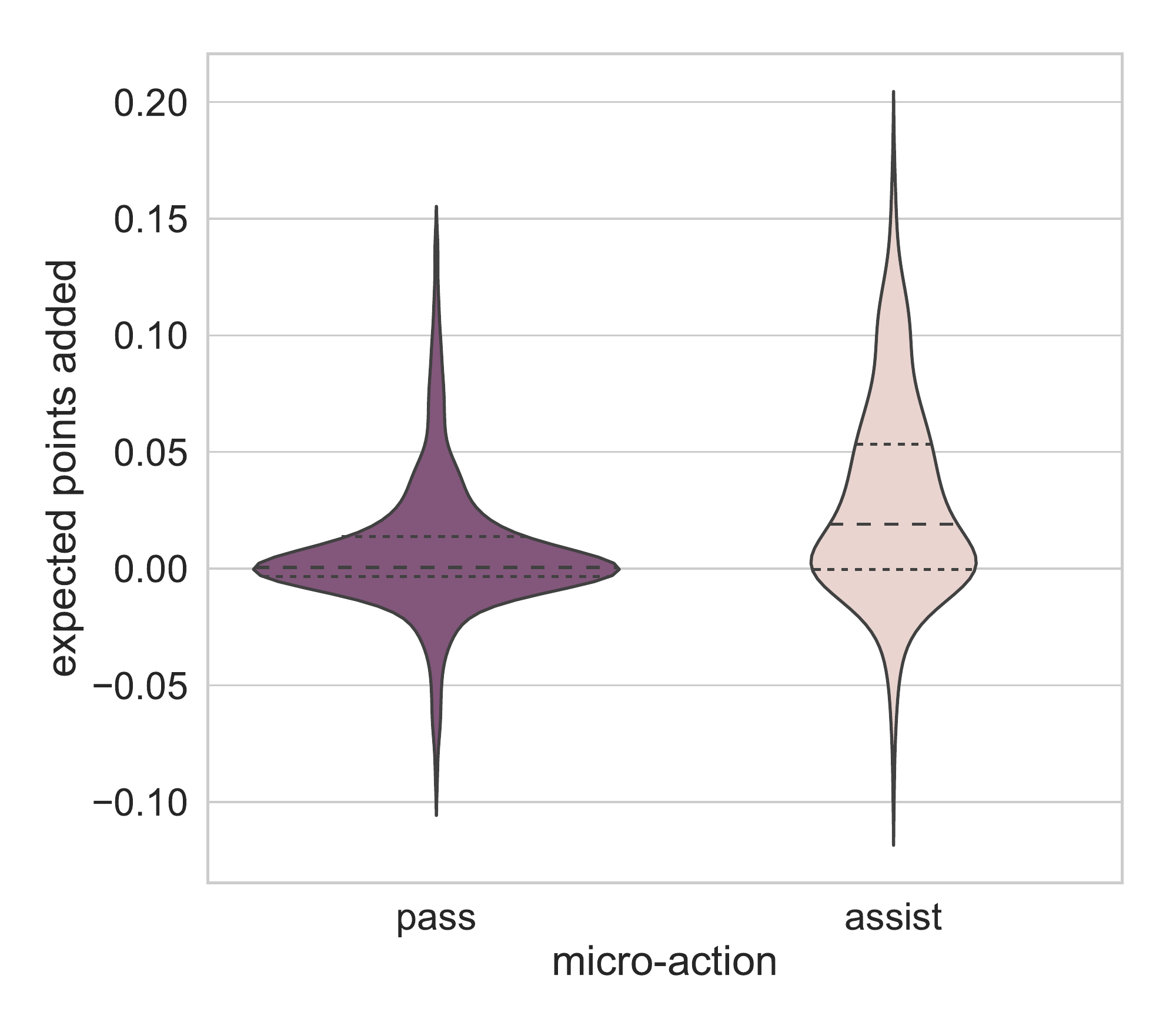}
\caption{Violin Plot (displays rotated kernel density estimation) of expected points added  for randomly sampled passes.}
\label{fig:micro_act}
\end{figure}
\begin{figure}[ht]
\centering
\includegraphics[width=\linewidth]{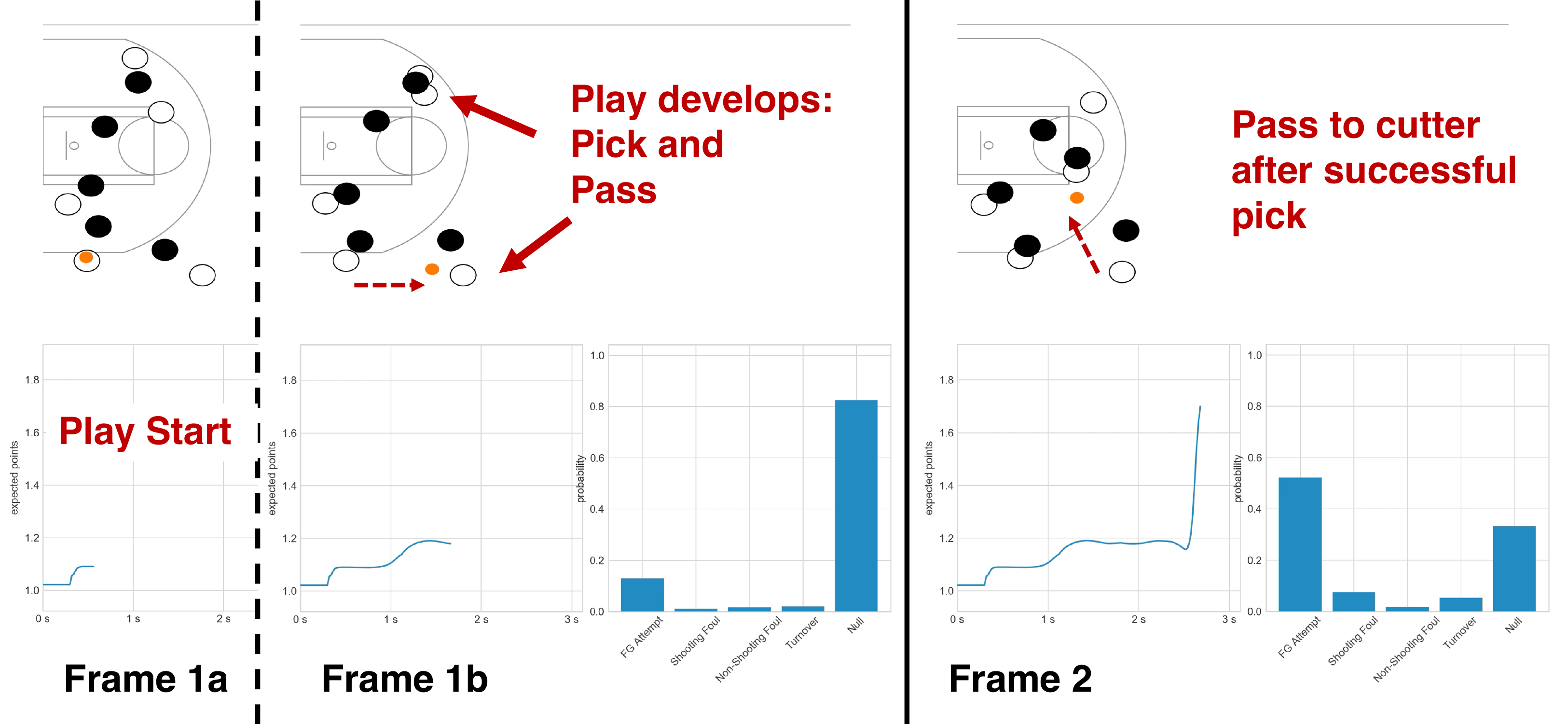}
\caption{Snapshots of play with running value of expected points and terminal-action probability estimates at each moment. Frame 1a shows the initial moments of the play. Frame 1b shows the play developing: Klay Thompson receives a screen and Patrick McCaw receives the ball; {\method} slightly increases probability of a field foal attempt. Frame 2 shows Thompson receiving the ball just before taking a shot; {\method} greatly increases probability of a field goal attempt, hence the expected points increases.}
\label{fig:snap152}
\vspace{-0.1in}
\end{figure}
\begin{figure*}[ht]
\centering
\includegraphics[width=\textwidth]{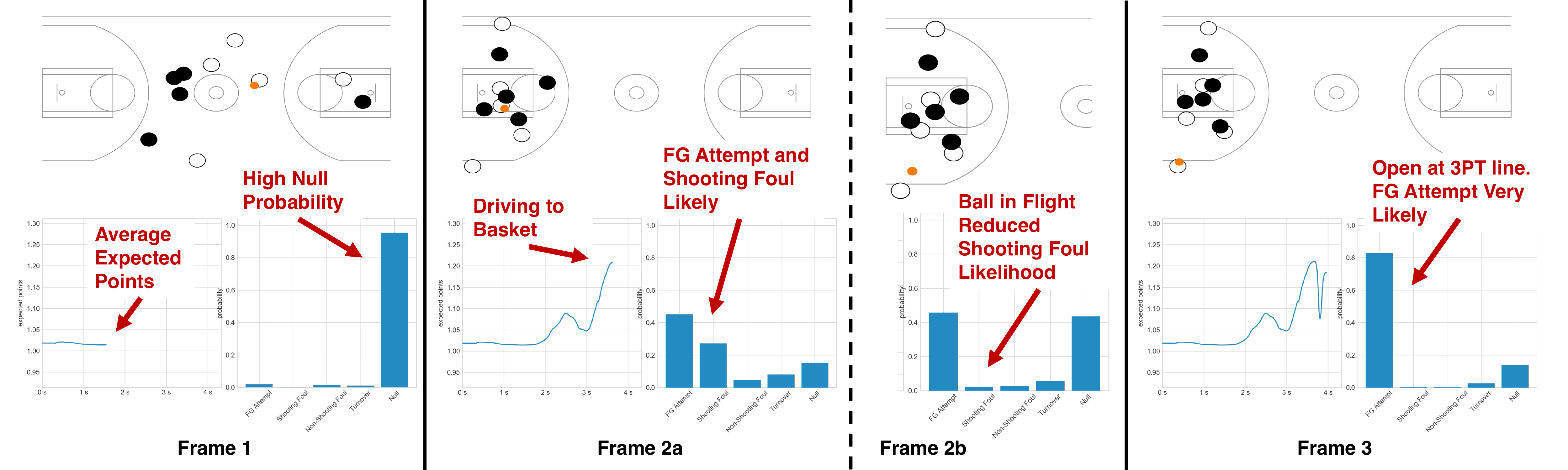}
\caption{Snapshots of play with running value of expected points and terminal-action probability estimates at each moment. Frame 1a shows average expected points (ball starts in backcourt for most plays). Frame 2a shows Durant surrounded by defenders in the key; {\method} increases likelihood of both field goal attempt and shooting foul. Frame 2b shows likelihood of shooting foul dropping as the ball is in flight. Frame 3 shows Ian Clark open in the corner, receiving the ball; {\method} maintains low likelihood of shooting foul, but increases likelihood of a shot attempt.}
\label{fig:snapshot170}
\end{figure*}%
%
\subsection{Evaluating Micro-Actions} 

\textbf{Evaluating Passes:} 
What is the difference in value between a standard pass and one which leads directly to a scoring event ? Using {\method} we can evaluate. 
In particular, we compute the expected points added (see Equation \ref{eqn:expectedpointsadded}) for 512 randomly sampled passes and 512 randomly sampled assists (that is passes that lead to a direct made FG). We take $\varepsilon = 21$ which corresponds to a 5 moment (fifth of a second) shift around the point of interest (when accounting for the blind-spot). Figure \ref{fig:micro_act} displays a Violin Plot and quartiles for the expected points added among the different types of passes. Assists are particularly important in basketball and are an indicator of good ball movement and team work. 
Hence, we should expect {\method} to value passes leading to a scoring event (assists) more highly then those that do not (e.g., a pass in the backcourt). 
This is clearly the case as can be seen by the estimated distributions given in Figure \ref{fig:micro_act}. 
Over 72\% of assists increase the expected points. 
In contrast, standard passes have high density near zero. 
Furthermore, a Kolmogorov-Smirnov test between the two distributions, rejects the null hypothesis that the two distributions are equal  at the 0.1\% level. 

\textbf{Realtime Application:} Figures \ref{fig:snap152} and \ref{fig:snapshot170} demonstrate snapshots of a realtime application of {\method} with videos available at \href{https://github.com/anthonysicilia/DeepHoopsRealtimeApplication}{https://github.com/anthonysicilia/DeepHoopsRealtimeApplication}. 

In Figure \ref{fig:snap152}, Klay Thompson cuts down the lane after a receiving a screen. Patrick McCaw passes to Thompson and he makes a turnaround fade away jump shot from 19 feet out. {\method} recognizes the play beginning to develop as probability estimates of the terminating action are shifted away from {\none} and toward field goal attempt, causing a slight increase in the expected points (Frame 1a to 1b). Then, after a successful screen, with Thompson cutting toward the left elbow, {\method} substantially increases the probability of a field goal attempt, consequently estimating a much higher expected points value (Frame 2). 

In Figure \ref{fig:snapshot170}, Kevin Durant drives towards the basket, but then dishes the ball to Ian Clark who makes a three point jump shot from 25 feet out. This play demonstrates {\method}' ability to identify when a shooting foul is likely and when it is not. As Durant approaches a group of defenders in the key, the likelihood of both a field goal attempt and a shooting foul increase, while the likelihood of {\none} goes down (Frame 1 to Frame 2a). During Durant's pass to an open Ian Clark on the left wing, the probability of a shooting foul drops significantly, as the ball is in flight, but the likelihood of a shot remains relatively stagnant (Frame 2b). When Clark receives the ball moments later, the probability of a field goal attempt increases substantially and the probability of a shooting foul remains low as he has no nearby defenders (Frame 3).

%% file: discussion.tex
\section{Conclusions and Discussion}
\label{sec:discussion}

In this paper we introduce a deep learning framework, {\method}, that is able to track the expected points to be scored by the offense during a possession in a basketball game. 
{\method} takes into consideration the players that are on the court during the possession as well as their spatio-temporal distribution. 
Our evaluations indicate that {\method} exhibits well calibrated estimates for the probability distribution of the possession outcomes. 
We further showcase how {\method} can be used to evaluate micro-actions that have traditionally been challenging to evaluate (e.g., how a standard pass differs from an assist). 

In the current implementation, for function $\nu$ (Figure \ref{fig:nu}) defined to assign the value above expectation, we have used league-average statistics. This can be easily adjusted to each lineup that is on the court for the possession examined. 
Furthermore, as part of our future work we are going to explore different approaches to player embedding (e.g., an offline embedding using boxscore statistics for players). Additionally, we will also explore the application of this deep learning framework to estimating the expected points during play in other sports (and in particular American football). 

{\bf Acknowledgements:} We would like to thank Dan Cervone for his valuable comments on our work.